\documentclass[aps,prl,twocolumn,groupedaddress,superscriptaddress,reprint,showpacs]{revtex4-2}
\usepackage{graphicx} 
\usepackage{tikz}
\usepackage{amsmath}
\usepackage{amssymb}
\usepackage{color}
\usepackage{float}
\usepackage{epstopdf}
\usepackage[normalem]{ulem}
\usepackage{pgfplots,pgfplotstable}
\usepackage{soul}
\usepackage{accents}
\usepackage{nameref}
\usepackage{balance}
\usepackage{upgreek}
\usepackage[all]{nowidow}
\usepackage{comment}
\usepackage{dsfont}
\usepackage{cleveref}
\usepackage{array} 				
\usepackage{booktabs} 			
\usepackage{ctable}

\begin{document}

\title{Mechanical activity and odd elasticity of passive, 2D chiral metamaterials}


\author{Mohamed Shaat}
\author{Harold S. Park}
\email[]{parkhs@bu.edu}
\affiliation{Department of Mechanical Engineering, Boston University, Boston, MA 02215, USA}


\date{\today}

\begin{abstract}
We demonstrate a general route to making active, odd elastic solids from passive chiral elements that can act as sources of mechanical work by violating static equilibrium without internal sources of energy or momentum.  We further demonstrate that by starting from a discrete, Newtonian mechanics viewpoint of the chiral unit cell, we can develop the continuum field equations for isotropic 2D chiral metamaterials that reveal odd elasticity, while elucidating the structure-property relationships underpinning the chiral, elastic moduli that enable the mechanical activity of the chiral metamaterials.  By demonstrating the energy gain of chiral metamaterials as they undergo quasistatic deformation cycles, we show a new route to designing active, non-reciprocal mechanical metamaterials that can operate at zero frequency.
\end{abstract}


\maketitle


A passive solid operating under thermodynamic equilibrium interacts with its surroundings through reversible processes, and thus its deformations must be path-independent, which requires knowledge of only the initial and final equilibrium states~\cite{De-Groot1962}. Deviation from this passive behavior would require an active solid that is capable of violating energy conservation either by irreversible interactions with its surrounding environment, or through the motion or activity of internal mechanisms or energy sources. Therefore, there has been significant interest in designing and obtaining active solid metamaterials, which can violate thermodynamic equilibrium, and thus do work on their surroundings.

Various routes have been established to make active solid metamaterials that can achieve thermodynamic non-equilibrium~\cite{Das2020,Shankar2020,Asadchy2020,Nassar2020}. Most of these routes involve the use of space-time modulators to break time reversal symmetry~\cite{Asadchy2020,Taravati2017,Correas-Serrano2016,Hadad2015,Shaltout2015,Estep2014,Fang2012a,Fang2012,Trainiti2019,Huang2019,Lu2020}, dynamics to achieve activity and non-reciprocity~\cite{Trainiti2019,Huang2019,Nash2015,Zhao2020}, mechanisms of feedback/control to generate linear or angular momentum~\cite{Rosa2020,Sirota2020,Brandenbourger2019}, or mechanisms of local sensing, actuation and control  to create motion, generate work, and violate thermodynamic equilibrium~\cite{Chen2021}. These routes have led to the development of active robotic metamaterials~\cite{Brandenbourger2019}, gyroscopic topological metamaterials~\cite{Nash2015}, and metamaterials for realizing non-Hermitian~\cite{Li2019,Scheibner2020,Ghatak2019,Gong2018,Li2019a,Kunst2019,Yao2018} and non-reciprocal~\cite{Taravati2017,Correas-Serrano2016,Hadad2015,Shaltout2015,Estep2014,Sounas2013,Jin2019,Coulais2017,He2018} physical systems.

Recently, the concept of odd elasticity was introduced to demonstrate a new class of solids that can be active through quasistatic deformations~\cite{Scheibner2020}. Specifically, odd elastic solids have a non-symmetric elasticity tensor, which implies they do not conserve angular momentum and energy due to the generation of an internal, non-conservative torque from the spinning of an active element. Odd elastic solids have been proposed theoretically based on active, non-reciprocal springs~\cite{Scheibner2020}, and experimentally based on active elements of metabeams with piezoelectric patches~\cite{Chen2021}. Whereas these previously proposed odd elastic solids can be active at zero-frequency, they required the presence of internal energy sources such as batteries~\cite{Scheibner2020}, or control systems involving electrical circuitry~\cite{Chen2021} to actuate and control the motion of the utilized active components. An important question thus arises, which we resolve here -- given this need for \textit{active elements} with internal sources of energy and momentum to obtain active metamaterials, can an odd elastic, active metamaterial be developed using only \textit{passive elements} that do not require internal sources of energy or momentum to operate? Furthermore, and of equal importance, because current manifestations of odd elastic solids begin assuming the odd elastic constitutive relationship~\cite{Chen2021,Scheibner2020,Scheibner2020a,Braverman2021}, it is not clear how the constitutive equations and continuum field equations that describe its observed odd elastic, and thus mechanically active behavior, originate from the microstructure, material properties, and kinematical fields of the solid.

Here, we demonstrate an isotropic 2D metamaterial made of passive, chiral elements that is mechanically active, and acts as a source of mechanical work through quasistatic deformations.  We first show that the metamaterial gains its mechanical activity through a non-conservative twist of its passive, chiral elements that is spontaneously generated upon deforming the solid elastically, which results in both energy and angular momentum not being conserved.  The non-conservative twist further enables the emergence of odd elasticity and violation of static thermodynamic equilibrium, which introduces a new class of odd elastic, active matter that can be active without internal sources of energy or momentum, circumventing the aforementioned limitations of the existing active metamaterials~\cite{Trainiti2019,Nash2015,Rosa2020,Sirota2020,Brandenbourger2019,Chen2021,Li2019,Scheibner2020}. We next demonstrate based on the mechanics of the discrete structure of the chiral element, the formalism of the continuum field equations that can capture the odd elastic behavior of chiral metamaterials.  Finally, we demonstrate the ability of chiral metamaterials to do work on their surroundings though no energy is dissipated by their passive, chiral elements.

The specific structure of the considered chiral metamaterial in Fig.1(a) is a 2D lattice of chiral elements, i.e. a central rigid circle and four elastic ligaments (Fig.1(b)). The rigid circle possesses three kinematical degrees of freedom, i.e. two displacements $u_x$ and $u_y$ and rotation about its center ${\theta }_z$, and the ligaments are elastic beams that can stretch and bend, with elastic modulus $E$, area moment of inertia $I$, and cross-sectional area $A$, and thus the chiral element is comprised only of passive elements. Two parameters define the geometry of the chiral metamaterial; the chiral angle ${\alpha }_0$ and the lattice length $a$, where the radius of the circle $R=(a/2)\mathrm{sin}{\alpha }_0$, and the length of the ligament $L=a\mathrm{cos}{\alpha }_0$ (Fig.1(b)).

\textit{Non-conservative Twist of Chiral Elements.--} Previously proposed systems that enable odd elasticity depended on internal energy sources to generate the non-conservative torques that enable non-conservation of angular momentum and energy~\cite{Scheibner2020,Chen2021,Fruchart2021}.  Here, we demonstrate the manifestation of the microscopic, non-conservative torque - that is required to enable odd elasticity - through a non-conservative, local twist ($\psi$), which is a kinematical field of the chiral element of the metamaterial that is developed when the ligaments of the metamaterial are elastically deformed, with no need for internal sources of energy or momentum control.

\begin{figure}
\includegraphics[width=1.0\linewidth]{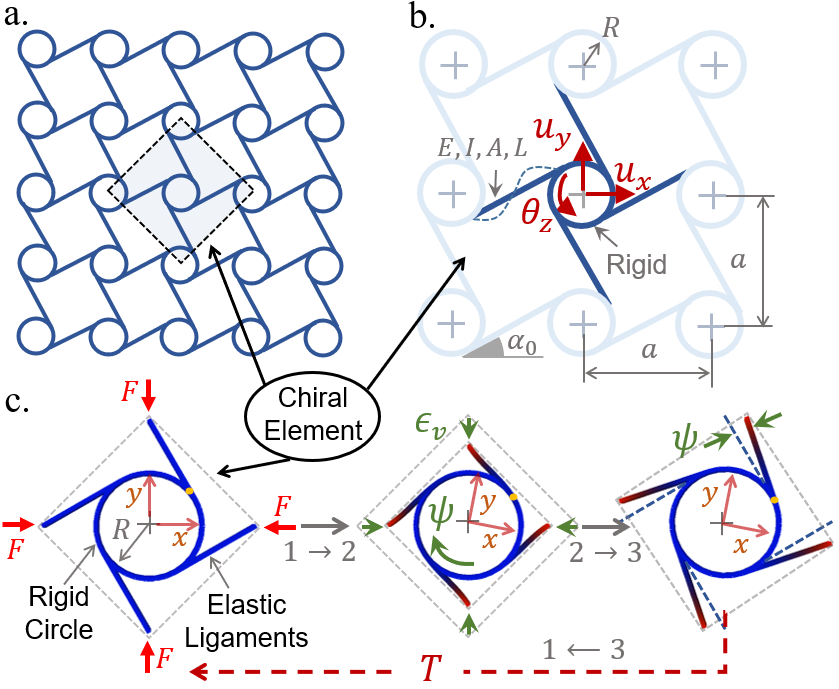}
\caption  {(a) Isotropic 2D metamaterial made of passive, chiral elements that is mechanically active and odd elastic. (b) Chiral element, which consists of a central rigid circle with three degrees of freedom, $u_x$, $u_y$, and $\theta_z$, and four solid, elastic ligaments that can stretch and bend.  (c) Finite element simulation demonstrating non-conservative twist $\psi$ of the chiral element.  In process 1$\rightarrow$2, the unit cell was subjected to compressive forces of $F=2$ N that produce a volumetric strain ${\epsilon }_v$, while a twist is spontaneously developed due to the bending and the chiral distribution of the ligaments around the rigid circle (middle). In process 2$\rightarrow$3, the forces were removed, which results in the volumetric strain being restored, though a permanent (non-conservative) twist $\psi$ remains (right), which can be recovered by an external torque $T=4FR\mathrm{cos}{\alpha }_0$ through the process 3$\rightarrow$1, indicating that the chiral element does not conserve both energy and angular momentum.}
\end{figure} 

Fig.1(c) shows finite element simulations~\cite{ABAQUS} of the passive, chiral element from Fig.1(b) undergoing a quasistatic loading-unloading deformation cycle. Under compressive forces $F$ in process 1$\rightarrow$2, the ligaments are bent and wrapped around the rigid circle, and thus a dilatation (volumetric strain) $\epsilon_v$ of the chiral element is produced, while a twist $\psi$ is developed at the center of the chiral element due to the bending and chiral distribution of the elastic ligaments around the rigid circle.  This twist is distinct from the rigid-body rotation of the entire chiral element $\theta_z$, as it is contingent on the ligaments being elastic and having a chiral distribution around the rigid circle. Thus, if the element is achiral ($\alpha_{0}=0$) or the ligaments are rigid (i.e., $\epsilon_v=0$), no twist is produced. When the chiral element is unloaded (process 2$\rightarrow$3), the dilatation is recovered as the ligaments are straightened, but the produced twist is not restored, demonstrating a non-conservative twist of the chiral element after the applied forces are removed, which thus achieves non-conservation of angular momentum without any internal energy or momentum sources (see supplemental video 1~\cite{SIVPRL}). The work done on the chiral element via the compressive forces during the loading process is fully recovered in the unloading process by the elastic ligaments. Nonetheless, because the twist is not conserved, additional work in the form of a torque $T=4FR \mathrm{cos}{\alpha }_0$ is required through the process 3$\rightarrow$1 to bring the chiral element back to its initial configuration, which demonstrates that the chiral element also achieves non-conservation of energy. See SI - S4.1~\cite{SIPRL} for additional details on the finite element simulations.

\textit{Odd Elasticity of 2D Chiral Metamaterials.--}  The concept of odd elasticity has been introduced to describe solids that reveal non-symmetric elasticity tensors, and it has been demonstrated for solids with active internal components~\cite{Scheibner2020}. Here, we demonstrate the odd elasticity of the chiral metamaterial shown in Fig.1(a), while introducing a new route to odd elasticity enabled by passive, chiral structures that can develop non-conservative twists.

Because mechanically active systems do not conserve energy, and because odd elastic effects cannot arise from a strain energy function~\cite{Scheibner2020}, we developed a discrete model of the chiral element highlighted in Fig.1(b) based on Newtonian mechanics, from which we determined the equivalent, linear elastic continuum model of the chiral metamaterial.  Specifically, we considered each circle in the chiral metamaterial as a rigid particle that possesses three kinematical degrees of freedom, i.e., $u_x$, $u_y$ and ${\theta }_z$, while the ligaments are elastic beams, where the interaction forces and couples between circles were determined in terms of the axial, bending, bending-torsion coupling, and torsional rigidities of the ligaments. We then balanced the forces along $x$- and $y$-directions and the moments about the $z$-direction to obtain three (discrete) equations of motion, which depend on the geometry and material properties of the ligaments, and the chiral angle $\alpha_{0}$. Then, by implementing a Taylor expansion, we determined the equations of motion at the continuum limit, where three continuum field equations of the chiral metamaterial were determined in terms of the kinematical degrees of freedom $u_x$, $u_y$ and ${\theta }_z$ (see SI -- S1~\cite{SIPRL}). The field equations were then simplified to the following two equations for $u_x$ and $u_y$, after the proper substitution for ${\theta }_z$ (see SI -- S2~\cite{SIPRL}):
\begin{eqnarray*}
\bar{M}u_{x,xx}+\left(\bar{\lambda}+\bar{\mu}\right)u_{y,xy}+\bar{\mu}u_{x,yy}+Q^{O}u_{y,yy}\\
+A^{O}u_{x,xy}+K^{O}u_{y,xx}=0~~~~(1)
\end{eqnarray*}
\begin{eqnarray*}
\bar{M}u_{y,yy}+\left(\bar{\lambda}+\bar{\mu}\right)u_{x,xy}+\bar{\mu}u_{y,xx}-Q^{O}u_{x,xx}\\
-A^{O}u_{y,xy}-K^{O}u_{x,yy}=0~~~~(2)
\end{eqnarray*}
\noindent where $()_{,\alpha}=\partial{()}/\partial{\alpha}$.  In addition, based on Eqs.(1) and (2), an asymmetric stress tensor was obtained such that ${\sigma}^{\alpha}=C^{\alpha \beta}u^{\beta}$, where ${\sigma}^{\alpha}$ and $u^{\beta}$ are, respectively, the stress and displacement gradient vectors based on Scheibner's notation~\cite{Scheibner2020}, and the elasticity tensor $C^{\alpha\beta}$ was determined to be:
\[{C^{\alpha\beta}}=\left[\begin{array}{cccc}
\widehat{B}+\overline{\alpha} & 0 & 0 & 0 \\ 
{\widehat{A}}^O & 0 & 0 & 0 \\ 
0 & 0 & \widehat{\mu }+\widehat{\gamma } & \widehat{\beta }+{\widehat{K}}^O \\ 
0 & 0 & \widehat{\beta }-{\widehat{K}}^O & \widehat{\mu }-\widehat{\gamma } \end{array}
\right] ~~~~~~~(3)\]
where $\widehat{B}=2(\overline{\lambda}+\overline{\mu})$,$\bar{\alpha }=\bar{M}-\left(\bar{\lambda}+2\bar{\mu}\right)$, $\widehat{\mu }={\overline{M}/2}-{\overline{\lambda }/2}+\overline{\mu }$, $\widehat{\gamma }={\overline{M}/2}-{\overline{\lambda }/2}-\overline{\mu }$, ${\widehat{A}}^O=A^O-K^O+Q^O$, ${\widehat{K}}^O=-{A^O}/2+{Q^O}/2+{3K^O}/2$, and $\widehat{\beta }={A^O}/2-{Q^O}/2+{K^O}/2$.

The elasticity tensor $C^{\alpha\beta}$ indicates that our 2D isotropic chiral metamaterial is odd elastic, as it can be decomposed into \textit{symmetric/even} ($C^{\alpha\beta}_{E}$) and \textit{skew-symmetric/odd} ($C^{\alpha\beta}_{O}$) elasticity tensors. The odd elastic tensor $C^{\alpha\beta}_{O}$ depends on the chiral, odd elastic moduli $\widehat{K}^O$ and $\widehat{A}^O$ that enable an odd elasticity of the chiral metamaterial in which both angular momentum and energy are not conserved (see SI -- S2~\cite{SIPRL}). The odd elastic behavior of the chiral metamaterial emerges only because of the chirality, as the moduli $\widehat{K}^O$ and $\widehat{A}^O$ both are zero if and only if the isotropic metamaterial is achiral (i.e., ${\alpha }_0=0$, see Fig.S3 in~\cite{SIPRL}), which demonstrates that the attainment of odd elasticity using passive elements is contingent on the elements being chiral. Physically, the chiral modulus $\widehat{A}^O$ enables dilatational deformations when the chiral metamaterial is subjected to internal torques, and local twists when the metamaterial is subject to applied pressure, while the chiral modulus $\widehat{K}^O$ enables shear deformations due to normal stresses and normal strains due to shear stresses.

It was recently noted that all 2D isotropic, odd elastic solids are chiral~\cite{Scheibner2020a}. However, the converse does not hold, as the 2D isotropic, chiral solid may not reveal odd elasticity. For instance, when $\widehat{A}^O=0$ and $\widehat{K}^O=0$, the elasticity tensor $C^{\alpha\beta}$ is symmetric (i.e., energy and momenta are conserved) though it is chiral when $\widehat{\beta}\neq 0$, and hence the 2D isotropic, chiral metamaterial is not odd elastic. Nonetheless, a 2D isotropic, chiral metamaterial with $\widehat{\beta }=0$ (i.e., $Q^O=K^O+A^O$) is always odd elastic, as the chirality requires having either $\widehat{A}^O$ or $\widehat{K}^O$ non-zero.

Beyond the odd elastic tensor $C^{\alpha\beta}_{O}$, the elasticity of the chiral metamaterial still differs from the classical, Cauchy elasticity in having the symmetric tensor $C^{\alpha\beta}_{E}$ depends on the non-classical moduli $\widehat{\beta}$, $\widehat{\alpha}$, and $\widehat{\gamma}$, along with the classical bulk modulus $\widehat{B}$ and shear modulus $\widehat{\mu}$. According to Eq.(3), the chiral modulus $\widehat{\beta}$ enables the metamaterial to exhibit chiral deformations where the axial stress produces shear and the shear stress produces normal strain, even if the metamaterial is not odd elastic. Nonetheless, these chiral deformations are fully recovered by means of energy conservation, and the chiral metamaterial is even elastic when $\widehat{A}^O=0$ and $\widehat{K}^O=0$. On the other hand, $\widehat{\alpha}$ and $\widehat{\gamma}$ are achiral, constrained moduli that result in additional axial and shear strains when the metamaterial is subjected to axial and shear stresses, respectively. These moduli are obtained in the field equations because the base structure of the metamaterial is a grid that results in  the displacement fields $u_x$ and $u_y$ being uncoupled when $\alpha_0=0$, where according to Eqs.(1) and (2), $\bar{\lambda}$+$\bar{\mu}$=0 and $A^O$, $K^O$, and $Q^O$ are zero when $\alpha_0=0$.  

\begin{figure*}
\includegraphics[width=1.0\linewidth]{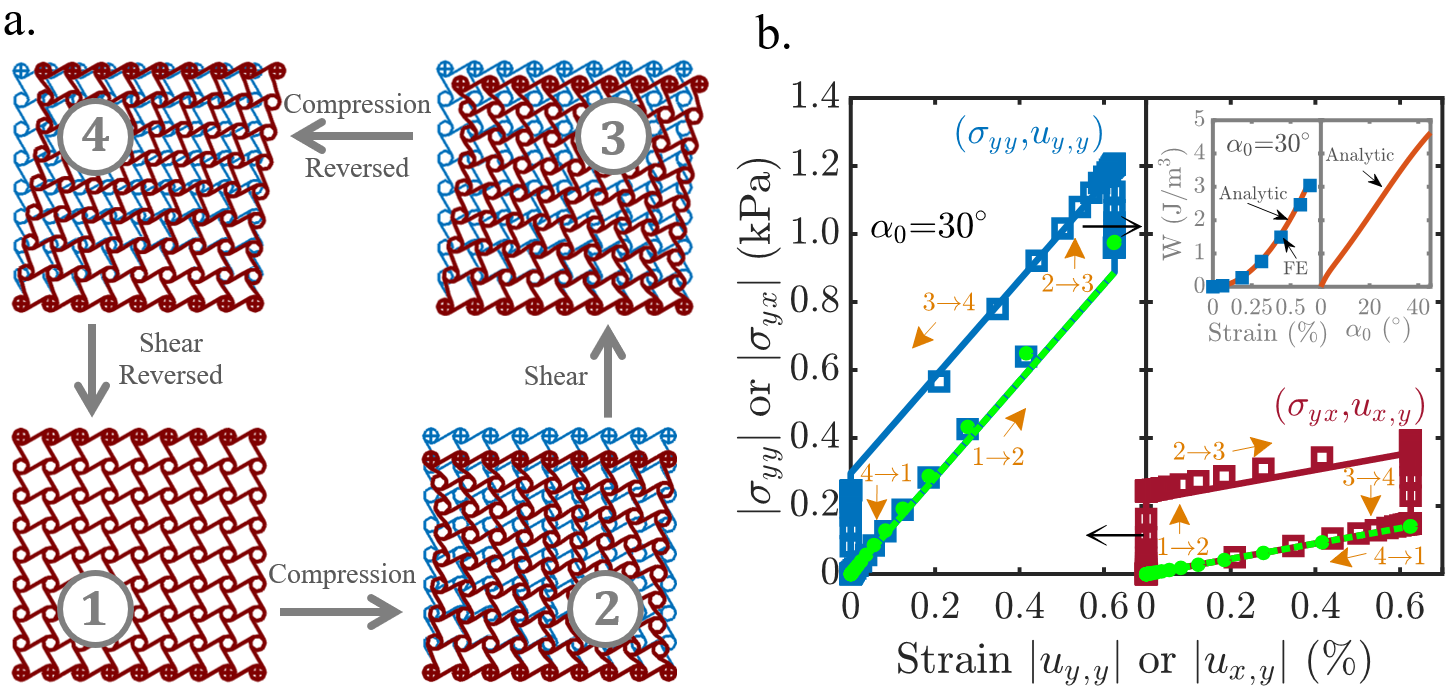}
\caption  {Mechanical activity of chiral metamaterials. (a) A chiral metamaterial undergoes a strain-controlled, quasistatic deformation cycle of a sequence of compressive ($u_{y,y}$) and shear ($u_{x,y}$) strains (see SI - S3.3 and S4.2 for more details~\cite{SIPRL}). (b) The axial stress $\sigma_{y,y}$ and the shear stress $\sigma_{y,x}$ versus the applied strain through the deformation cycle. The cycle represented by the solid lines is the predictions of the analytical model based on Eq.(3), while the cycle represented by the squares is the predictions of the finite element simulations. The chiral metamaterial can do work on its surroundings at two instances indicated by black arrows through the cycle. The insets show the odd elastic work per unit volume $W$ developed by the chiral metamaterial during the cycle versus the applied strain (left) and the chiral angle $\alpha_0$ (right). The green dashed lines and the green circles represent, respectively, the analytic and the finite element results of a non-chiral, bulk material undergoing the same deformation cycle.}
\end{figure*} 

Previous works have used micropolar elasticity to model the mechanics of chiral solid materials~\cite{Liu2012,Joumaa2011,Lakes1982,Lakes2001,Natroshvili2006,Frenzel2017,Frenzel2019,Chen2020,Chen2021a}. However, because micropolar elasticity is based on a free energy function and thus enforces energy conservation and conservation of linear and angular momenta, the odd, chiral moduli $\widehat{K}^O$ and $\widehat{A}^O$ are not present in micropolar elasticity,  and thus the odd elasticity described above along with the mechanical activity of chiral metamaterials cannot be captured by micropolar elasticity (see SI -- S5~\cite{SIPRL}). Revealing the odd elasticity of chiral metamaterials instead requires obtaining the continuum field equations based on the mechanics of the discrete structure of the chiral element, as presented here, which enables a clear understanding of how the constitutive equations and the odd elastic moduli originate from the microstructure and material properties of the solid. As shown in SI -- Eq.(S25) and Fig.S3~\cite{SIPRL}, we determined the material moduli $\overline{M}$, $\overline{\lambda }$, $\overline{\mu }$, $K^O$, $A^O$ and $Q^O$ as functions of the geometrical and material properties of the ligaments ($E,I,L,A$), as well as the chiral geometry (${\alpha }_0$ and $a$).

\textit{Mechanical Activity of 2D Chiral Metamaterials.--} As discussed previously, different routes have been established to make a solid active, most of which involve the use of active elements with internal sources of energy and momentum control. Here, we demonstrate that chiral metamaterials introduce a new route to making active solids from passive elements by elucidating their ability to perform mechanical work on their surroundings through a quasistatic deformation cycle without internal sources of energy or momentum.  

The activity and odd elasticity of our chiral metamaterial can be verified by determining the energy gain (i.e., work done by the chiral metamaterial) when it undergoes a quasistatic deformation cycle where the structural configuration is the same at the beginning and the end of the cycle. Therefore, we considered a chiral metamaterial that is formed by $8\times 8$ chiral elements with a chiral angle $\alpha_0=30^{\circ}$ and lattice length $a=20$ mm, where the circles are rigid and the ligaments are made of thermoplastic polyurethane (TPU, elastic modulus $E=150\ \mathrm{MPa}$) with 2 mm thickness and 10 mm width. The chiral metamaterial was subject to the strain-controlled, quasistatic deformation cycle shown in Fig.2(a), while the stresses and the energy gain during the cycle were determined by means of analytical solutions based on Eq.(3) and finite element simulations, as shown in Fig.2(b) (see SI - S3 and S4~\cite{SIPRL} for more details).  In the process 1$\rightarrow$2, a compressive strain $u_{y,y}$ is applied, and an axial stress $\sigma_{yy}=\hat{E}u_{y,y}$ is developed in the chiral metamaterial, while because of its odd elasticity, the chiral metamaterial also develops an additional odd shear stress $\sigma_{yx}^{O}$ that depends on the chiral, odd elastic modulus $Q^O$, where $\hat{E}=(\bar{M}(\bar{M}+\bar{\lambda})-\bar{\lambda}^2)/(\bar{M}+\bar{\lambda})$ is the effective Young's modulus of the chiral metamaterial. In the process 2$\rightarrow$3, a shear strain $u_{x,y} $ is then applied where the shear stress is increased to be $\sigma_{yx}=\bar{\mu}u_{x,y}+\sigma_{yx}^{O}$, while through the odd elasticity of the chiral metamaterial an additional odd axial stress $\sigma_{yy}^{O}$ that depends on the chiral, odd elastic modulus $K^{O}$ is also developed, where the overall axial stress in the metamaterial increases to be $\sigma_{yy}=\hat{E}u_{y,y}+\sigma_{yy}^{O}$, as shown in Fig.2(b).  Then, the chiral metamaterial is unloaded where the compression is reversed in the process 3$\rightarrow$4 and the shear is reversed in the process 4$\rightarrow$1, while the axial and shear stresses in the metamaterial decrease to become zero at the end of the cycle (Fig.2(b)). 

The calculations of the strain energy density (i.e. the area under the stress-strain curves) indicate that the chiral metamaterial can act on its surroundings with odd elastic work at two instances while it undergoes the deformation sequence shown in Fig.2(a) (see SI-S3.3 and S4.2~\cite{SIPRL} for details on energy calculations). Through the process 1$\rightarrow$2, the chiral metamaterial can act on its surroundings by an odd elastic work per unit volume of $(1/2)\sigma_{yx}^{O}u_{x,y}$, while it can also act on its surroundings with an odd elastic work per unit volume of $(1/2)\sigma_{yy}^{O}u_{y,y}$ through the process 2$\rightarrow$3, and thus the total odd elastic work per unit volume (i.e., energy gain) of the chiral metamaterial when it undergoes the deformation cycle shown in Fig.2(a) becomes $W=\sigma_{yx}^{O}u_{x,y}+\sigma_{yy}^{O}u_{y,y}$, which increases as the applied strain through the cycle increases (see the inset in Fig.2(b)). To verify that this behavior is unique for chiral metamaterials, we considered an achiral, bulk material through the same strain-controlled deformation cycle, and the stresses and the strain energy density developed during the cycle were determined (see the green dashed-lines and circles in Fig.2(b)). The achiral, bulk material traced the same stress-strain curve while loading and unloading, where no energy gain was observed during the cycle, and thus the material cannot do work on its surroundings. This demonstrates that the ability of the metamaterial to deliver work is contingent on the use of chiral elements, where the odd elastic work $W$ is zero if it has achiral elements (i.e., if $\alpha_{0}= 0$, $\sigma_{y,y}^{O}=0$, $\sigma_{y,x}^{O}=0$, and $W = 0$, see the inset in Fig.2(b)).

\textit{Conclusions.} -- We presented a general route to making active, odd elastic solids based on passive elements that do not require internal sources of energy or momentum.  Specifically, embedding chirality into the solid microstructure enables activating 
non-conservative kinematical fields that can achieve angular momentum and energy non-conservation, and hence odd elasticity, thus enabling it to do work on its surroundings during a quasistatic deformation cycle. We anticipate that the relationships between chirality, odd elasticity and mechanical activity we derived will enable many new roads for investigation, including the use of passive chiral metamaterials for the realization of novel non-Hermitian and non-reciprocal physical systems, embedding novel functionality into mechanical metamaterials, and developing new field theories to understand, analyze and design future odd elastic systems and structures.

\begin{acknowledgments}
The authors acknowledge the support of ARO grant W911NF-21-2-0091.
\end{acknowledgments}

\providecommand{\noopsort}[1]{}\providecommand{\singleletter}[1]{#1}%

\end{document}